\documentclass[twocolumn,prl,twoside,showpacs,preprintnumbers,floatfix]{revtex4}

\usepackage{amsmath}
\usepackage{amssymb}
\usepackage{graphicx}
\usepackage{dcolumn}
\usepackage{bm}

\def\m#1{\mathrm{#1}}
\def\Eq#1{(\ref{eq:#1})}
\def\d{\mathrm{d}}

\def\epsilon{\varepsilon}
\def\theta{\vartheta}
\def\rho{\varrho}

\begin{document}


\title{Liquid-liquid interfacial tension of electrolyte solutions}

\author{Markus Bier}
\email{m.bier@uu.nl}

\author{Jos Zwanikken}

\author{Ren\'{e} van Roij}

\affiliation
{
   Institute for Theoretical Physics, 
   Utrecht University, 
   Leuvenlaan 4, 
   3584CE Utrecht, 
   The Netherlands
}

\date{June 12, 2008}

\begin{abstract}
It is theoretically shown that the excess liquid-liquid interfacial tension between two electrolyte solutions 
as a function of the ionic strength $I$ behaves asymptotically as $\mathcal{O}(-\sqrt{I})$ for small $I$ and 
as $\mathcal{O}(\pm I)$ for large $I$. The former regime is dominated by the electrostatic potential due to an 
\emph{unequal partitioning} of ions between the two liquids whereas the latter regime is related to a finite 
interfacial thickness. The crossover between the two asymptotic regimes depends sensitively on material parameters
suggesting that, depending on the actual system under investigation, the experimentally accessible range of
ionic strengths can correspond to either the small or the large ionic strength regime. In the limiting case
of a liquid-gas surface where ion partitioning is absent, the image chage interaction can dominate the surface 
tension for small ionic strength $I$ such that an Onsager-Samaras limiting law $\mathcal{O}(-I\ln(I))$ is 
expected. The proposed picture is consistent with more elaborate models and published measurements.
\end{abstract}

\pacs{68.05.-n, 82.45.Gj, 68.03.Cd}

\maketitle


The temporal stability of liquid-liquid emulsions, which is of enormous importance for applications in, e.~g., 
chemical, pharmaceutical, food, and cosmetic industries, largely hinges on the liquid-liquid interfacial tension 
\cite{Binks2000} modified by surfactants, cosurfactants, and even colloidal particles \cite{Aveyard1985}. In order 
to theoretically understand and predict the liquid-liquid interfacial tension as a function of additives a first 
step is modeling a liquid-liquid interface in the presence of electrolytes but in the absence of surfactants. 
Remarkably, the dependence of the liquid-liquid interfacial tension on the electrolyte concentration is, in contrast 
to the liquid-gas surface tension \cite{Manciu2003}, not well understood. This is quite astonishing because
liquid-liquid interfaces have been investigated for a long time by means of electrocapillary measurements
\cite{Volkov1996}. The few reported measurements of the liquid-liquid interfacial tension as a function
of the ionic strength known to the authors, Refs.~\cite{Guest1939,Aveyard1976,Girault1983}, seem to confirm 
the linear relation at large ionic strengths well-known from liquid-gas surface tension measurements 
\cite{Weissenborn1996}. At low ionic strengths the liquid-gas surface tension exhibits the Jones-Ray 
effect, i.~e., a minimum of the surface tension as a function of the ionic strength \cite{Jones1935}, whose 
analog for liquid-liquid interfacial tensions has been addressed in the experimental literature, to the authors' 
knowledge, only in Ref.~\cite{Guest1939}. Theoretical approaches to liquid-gas surfaces are very often based on
the assumption that the gas phase is completely free of ions \cite{Manciu2003}, which leads to a charge neutral
liquid phase. Considering the image charge interaction as dominating the liquid-gas surface tension at low ionic 
strength the Onsager-Samaras limiting law can be derived \cite{Onsager1934}. However, assuming a
non-vanishing ionic strength in the gas phase, Nichols and Pratt found indications that the liquid-gas surface
tension in some instances can also scale with the square root of the ionic strength in the low salt limit 
\cite{Nichols1984}. By means of an elaborate Ginzburg-Landau-like model for liquid-liquid interfaces, taking ion 
densities and solvent composition explicitely into account, Onuki recently observed such a square root behavior for 
the liquid-liquid interfacial tension, too \cite{Onuki2006}. 
It is the aim of this letter to argue in terms of a \emph{minimal} model that unequal ion partitioning and charge 
separation are the key features of liquid-liquid interfaces of electrolyte solutions at low ionic strength.
Onsager-Samaras-like behavior can be found only in the absence of unequal ion partitioning and is therefore 
unexpected for liquid-liquid interfaces. 

In order to define the present model consider an infinite system composed of two homogeneous solvents $A$ and $B$ 
located within the half spaces $z<0$ and $z>0$, respectively, of a Cartesian coordinate system. In the interior 
of the solvents the relative dielectric constant $\epsilon(z)$ at position $z$ is given by 
$\epsilon(z<0)=\epsilon_A$ and $\epsilon(z>0)=\epsilon_B$. In the following the abbreviation 
$n:=\sqrt{\epsilon_A/\epsilon_B}$ will be useful. Monovalent ions are distributed 
in \emph{both} solvents giving rise to local equilibrium number densities $\rho_\alpha(z)$ at position $z$ with 
$\alpha=+$ and $\alpha=-$ denoting cations and anions, respectively. Deep in the solvent phases local charge 
neutrality holds, i.~e., $\rho_\alpha(-\infty)=:\rho_A$ and $\rho_\alpha(\infty)=:\rho_B$. The \emph{partition 
coefficient} is defined by $p:=\sqrt{\rho_A/\rho_B}$.
In general, the solubility of $\alpha$ ions differs in the two solvents. This effect can be described by 
solvent-induced potentials $V_\alpha(z)$ which take the limiting values $V_\alpha(-\infty):=0$ and 
$V_\alpha(\infty):=f_\alpha$ where $f_\alpha$ is the solvation free energy difference of an $\alpha$ ion in 
solvent $B$ as compared to solvent $A$. Verwey and Niessen \cite{Verwey1939} assumed the steplike form 
$V^\m{VN}_\alpha(z) = f_\alpha\Theta(z)$, where $\Theta$ denotes the Heaviside function. Such a model ignores 
interfacial effects due to an actually smooth dielectric function $\epsilon$, finite ion size, van der Waals forces,
solvation (structure making and structure breaking), and image charges \cite{Manciu2003}. All these effects depend
on material parameters of the system but they depend, with the exception of the image charge interaction, not 
directly on the ionic strength. Moreover, the image charge interaction decays as 
$\mathcal{O}(\exp(-2\kappa_{A,B}|z|)/|z|)$
with $\kappa_{A,B}^{-1}$ denoting the Debye screening length in phase $A$ for $z\rightarrow-\infty$ and
in phase $B$ for $z\rightarrow\infty$ \cite{Onsager1934,Manciu2003,Onuki2006}, whereas the electrostatic potential 
is expected to decay much slower as $\mathcal{O}(\exp(-\kappa_{A,B}|z|))$. Hence the image charge interaction is
expected to be negligible outside the interfacial region. A simple account of the mentioned interfacial effects 
is given by the \emph{shifted} Verwey-Niessen potentials $V_\alpha(z) := f_\alpha\Theta(z-s)$ where the discontinuity 
is located at position $z=s$, similar to the interface model by Johansson and Eriksson \cite{Johansson1974}.
Note that the electrostatic potential is the only interaction which is \emph{not} described by the solvent-induced 
potentials $V_\alpha$ because it is the longest-ranged ionic-strength-dependent interaction. Moreover, the shift
of the ion densities with respect to the solvent composition profile in Onuki's work \cite{Onuki2006} are compatible 
with the introduction of external fields similar to the present solvent-induced potentials $V_\alpha$. The 
location of the discontinuity of the solvent-induced 
potentials with respect to the dielectric interface at $z=0$ is a property of the solvents and the electrolyte. 
The analysis will in fact reveal that only changing the anion type can shift the discontinuity of $V_\alpha$ to 
the opposite side of the interface. Without restriction $s \geq 0$ is assumed, i.~e., solvent $B$ is \emph{defined} 
as the one where the discontinuity of $V_\alpha$ is located. 

The equilibrium structure represented by the density profiles $\rho_\alpha$ is most easily calculated in terms
of density functional theory \cite{Evans1979}. In units of the thermal energy $k_BT$, the elementary charge 
$e$, and the vacuum Bjerrum length $\displaystyle \ell=\frac{e^2}{4\pi\epsilon_\m{vac}k_BT}$ with the 
permeability of the vacuum $\epsilon_\m{vac}$, and within a mean-field theory ignoring ion-ion correlations, the
density functional of the grand potential per unit surface area 
\begin{eqnarray}
   \Omega[\rho_\pm] 
   & = & 
   \sum_{\alpha=\pm}\int\d z \rho_\alpha(z)\bigg(\ln(\rho_\alpha(z)) - 1 - \mu_\alpha 
   \nonumber\\
   & &
   \phantom{\sum_{\alpha=\pm}\int\d z}
   + V_\alpha(z) + \alpha\frac{1}{2}\phi(z,[\rho_\pm])\bigg)
   \label{eq:df}
\end{eqnarray}
is to be minimized with respect to $\rho_\alpha$. Here $\mu_\alpha$ is the chemical potential of species $\alpha$
and the electrostatic potential $\phi(z,[\rho_\pm])$ at position $z$, which is a functional of the ion density 
profiles $\rho_\pm$, fulfills the Poisson equation
\begin{equation}
   \frac{\d}{\d z}\epsilon(z)\frac{\d}{\d z}\phi(z,[\rho_\pm]) = -4\pi(\rho_+(z) - \rho_-(z))
   \label{eq:pe}
\end{equation} 
with the Dirichlet boundary conditions $\phi(-\infty) = 0$ and $\phi(\infty) = \phi_D$, where 
$\phi_D:=\frac{1}{2}(f_- - f_+)$ is the Donnan potential following from the local charge neutrality in the bulk
liquids. The electrostatic potential $\phi$ is continuous and it holds $\epsilon_A\phi'(0^-)=\epsilon_B\phi'(0^+)$, 
where a prime denotes a spatial derivative.
From the Euler-Lagrange equations corresponding to Eqs.~\Eq{df} and \Eq{pe} one readily derives for the 
\emph{shifted electrostatic potential} $\psi(z) := \phi(z) - \phi_D\Theta(z-s)$ the linearized 
Poisson-Boltzmann equation
\begin{equation}
   \frac{\d^2}{\d z^2}\psi(z) = \kappa(z)^2\psi(z), \quad z\not=0,s
   \label{eq:lpbe}
\end{equation}
with the homogeneous Dirichlet boundary conditions $\psi(\pm\infty)=0$ and the piecewise constant \emph{Debye
screening factor} $\kappa(z)$ defined by $\kappa(z)^2:=\kappa_A^2:=8\pi\rho_A/\epsilon_A$ for $z<0$, 
$\kappa(z)^2:=\kappa_i^2:=8\pi\rho_A/\epsilon_B$ for $z\in(0,s)$, and 
$\kappa(z)^2:=\kappa_B^2:=8\pi\rho_B/\epsilon_B$ for $z>s$. Moreover, the partition coefficient is 
found as $p=\exp((f_++f_-)/4)$.
The solution of Eq.~\Eq{lpbe} is
\begin{equation}
   \psi(z) = 
   \left\{\begin{array}{ll}       
      \displaystyle\frac{\phi_D}{D}\exp(\kappa_A z)                         & , z < 0     \\[5pt]
      \displaystyle\frac{\phi_D}{D}(\cosh(\kappa_i z) + n\sinh(\kappa_i z)) & , z\in(0,s) \\[5pt]
      \displaystyle-\frac{\phi_D}{D}\exp(-\kappa_B(z-s))                    &             \\[5pt]
      \times p(n\cosh(\kappa_i s) + \sinh(\kappa_i s))                      & , z > s
   \end{array}\right.
   \label{eq:sol}
\end{equation}
with $D := (1+np)\cosh(\kappa_i s) + (n+p)\sinh(\kappa_i s)$.
For the case $s=0$ the second line of Eq.~\Eq{sol} is empty.

If the Donnan potential does not vanish, $\phi_D\not=0$, a difference in solvation free energy leads to an
unequal partitioning of cations and anions on the two half spaces occupied by solvent $A$ and $B$. A measure 
for this unequal partitioning is the integrated charge density of the half space $z<0$
\begin{equation}
   \sigma_A 
   := 
   \int_{-\infty}^{0^-}\d z \sum_\alpha \alpha\rho_\alpha(z) 
   =
   -\frac{\phi_D\epsilon_A\kappa_A}{4\pi D}.
   \label{eq:sigmaAexact}
\end{equation}
For $\kappa_is \ll 1$ the integrated charge density $\sigma_A$ is constant to leading order in $\kappa_is$, i.~e., 
global quantities describing the ion partitioning within the Verwey-Niessen model ($s=0$) are not influenced by 
finite interface extensions smaller than the interfacial Debye length $\kappa_i^{-1}$. This finding a posteriori
justifies the application of the original Verwey-Niessen model to calculate droplet charges in 
Ref.~\cite{Zwanikken2007}.

The interfacial tension, however, is known to be highly sensitive to details of the interfacial structure.
In terms of the density functional $\Omega$ (see Eq.~\Eq{df}) the interfacial tension in excess of
the pure, salt-free liquid-liquid interface is given by 
$\Delta\gamma = \Omega[\rho_+,\rho_-]-\Omega[\rho_\m{ref},\rho_\m{ref}]$, where $\rho_\m{ref}$ is the steplike 
reference ion number density profile. For the excess interfacial tension with respect to the dielectric interface at 
$z=0$ the reference density is defined by $\rho_\m{ref}(z<0):=\rho_A$ and $\rho_\m{ref}(z>0):=\rho_B$ which
leads to
\begin{eqnarray}
   \Delta\gamma 
   & = & 
   2(1 - p^2)s\rho_B 
   \label{eq:gammaexact}
   \\
   & &
   - \frac{\phi_D^2\sqrt{\epsilon_B}p}{2\sqrt{2\pi}D}
   (n\cosh(\kappa_i s) + \sinh(\kappa_i s))\sqrt{\rho_B}.
   \nonumber
\end{eqnarray}
As the second term on the right-hand side of Eq.~\Eq{gammaexact} is of the order $\mathcal{O}(-\sqrt{\rho_B})$ 
for both $\rho_B\rightarrow 0$ and $\rho_B\rightarrow\infty$ one finds the following asymptotic behavior of
the excess interfacial tension:
\begin{equation}
   \Delta\gamma \simeq 
   \left\{
   \begin{array}{ll}
      \displaystyle
      -\frac{\phi_D^2\sqrt{\epsilon_B}}{2\sqrt{2\pi}}
      \frac{np}{1 + np}\sqrt{\rho_B}
      & ,\rho_B \rightarrow 0
      \\
      2(1 - p^2)s\rho_B 
      & ,\rho_B\rightarrow\infty.
   \end{array}
   \right.
   \label{eq:gammaasymp}
\end{equation}
As $n$ and $p$ are experimentally accessible, one can use Eq.~\Eq{gammaasymp} to determine $\phi_D$ or $s$.
The crossover, where the low-density asymptotics $\Delta\gamma=\mathcal{O}(-\sqrt{\rho_B})$ and the 
high-density asymptotics $\Delta\gamma=\mathcal{O}(\pm\rho_B)$ are of the same magnitude, takes place at 
the ionic strength
\begin{equation}
   \rho_B^\times := 
   \frac{\displaystyle \phi_D^4\epsilon_Ap^2}
        {\displaystyle 32\pi s^2(1 + np)^2(1 - p^2)^2}.
   \label{eq:gammacross}
\end{equation}
For $\rho_B > \rho_A$ one finds $\Delta\gamma(\rho_B\ll\rho_B^\times) < 0$ and 
$\Delta\gamma(\rho_B\gg\rho_B^\times) > 0$, i.~e., the excess interfacial tension vanishes near the crossover.
For $\rho_B < \rho_A$, on the other hand, $\Delta\gamma(\rho_B)<0$ for all $\rho_B$.
As the two bulk ion concentrations $\rho_A$ and $\rho_B$ are proportional to each other within the present model, 
one can choose either one calling it the \emph{ionic strength} $I$. Equation~\Eq{gammacross} leads to a 
corresponding crossover ionic strength $I^\times$. 

Equations~\Eq{gammaasymp} and \Eq{gammacross} are the main results of the present work which will be discussed in
the following.

The results presented so far have been derived from the \emph{linear} Poisson-Boltzmann equation Eq.~\Eq{lpbe}
which is expected to be reliable if $|\psi(z)| \ll 1$, i.e., $|\phi_D| \ll 1$. However, upon solving the 
\emph{non-linear} Poisson-Boltzmann equation derived from Eqs.~\Eq{df} and \Eq{pe} numerically, we found
the \emph{same} asymptotic dependence on the ionic strength $I$, 
$\Delta\gamma(I\ll \widetilde{I}^\times)=\mathcal{O}(-\sqrt{I})$ and 
$\Delta\gamma(I\gg \widetilde{I}^\times)=\mathcal{O}(\pm I)$, as in Eq.~\Eq{gammaasymp} with a crossover at 
$\widetilde{I}^\times \geq I^\times$ where the difference $\widetilde{I}^\times - I^\times$ increases with $|\phi_D|$.
Hence the asymptotic scaling of the interfacial tension difference $\Delta\gamma$ with the ionic 
strength $I$ and the existence of a crossover $I^\times$ are 
robust qualitative features of the linear theory when compared to the non-linear Poisson-Boltzmann theory. 
Moreover, by numerical fitting one obtains renormalized parameters $\phi_D^*$ and $s^*$ in Eq.~\Eq{gammaexact} 
such that $\Delta\gamma$ calculated within non-linear Poisson-Boltzmann theory is reproduced even 
\emph{quantitatively}.

As the asymptotic behavior of the excess interfacial tension $\Delta\gamma=\mathcal{O}(\pm I)$ for 
$I \gg I^\times$ in Eq.~\Eq{gammaasymp} involves the parameter $s$ one concludes that the finite size of the 
interfacial region is responsible for this asymptotics. This finding is confirmed by published measurements of 
liquid-liquid interfacial tensions \cite{Aveyard1976} and is in fact well-known from liquid-gas surface tensions 
\cite{Manciu2003,Weissenborn1996}. In contrast, the behavior $\Delta\gamma=\mathcal{O}(-\sqrt{I})$ for 
$I\ll I^\times$ in Eq.~\Eq{gammaasymp} can be attributed to the unequal ion partitioning because the prefactor of 
the asymptotics contains a term of electrostatic origin which vanishes if $\phi_D=0$. The latter regime, which
gives rise to a \emph{negative} contribution to the interfacial tension, is in contradiction to the 
Onsager-Samaras limiting law $\mathcal{O}(-I\ln(I))$ \cite{Onsager1934}, which contributes \emph{positively}
\cite{Nichols1984}. However, according to the present model, the image charge interaction is neglected in 
comparision to the electrostatic potential due to the unequal ion partitioning, whereas it is the dominating 
interaction within the Onsager-Samaras model \cite{Onsager1934}.
Therefore it can be concluded that unequal ion partitioning, which is expected to be a general phenomenon for 
liquid-liquid interfaces \cite{Onuki2006}, leads to $\Delta\gamma=\mathcal{O}(-\sqrt{I})$ for small $I$, whereas 
the absence of unequal ion partitioning gives rise to $\Delta\gamma=\mathcal{O}(-I\ln(I))$ \cite{Onsager1934}. The 
situation of a liquid-gas surface with non-vanishing ionic strength in the gas phase investigated by Nichols and Pratt 
\cite{Nichols1984} can be considered as the borderline between both scenarios such
that features of both, the square root and the Onsager-Samaras limiting law, can be visible.

From Eq.~\Eq{gammacross} one infers a high sensitivity of the crossover ionic strength $I^\times$
from the low ionic strength regime $\gamma(I \ll I^\times)=\mathcal{O}(-\sqrt{I})$ to the high ionic strength 
regime $\gamma(I\gg I^\times)=\mathcal{O}(\pm I)$ upon the model parameters $\phi_D$, $s$, and $p$, i.~e.,
upon the material parameters of the system. This obervation is also borne out by the results of Onuki 
\cite{Onuki2006}. Hence, depending on the actual system under investigation, $I^\times$ can be larger or smaller
than the experimentally available range of ionic strength as will be shown in the following. 

\begin{figure}[!t]
   \includegraphics[width=8.6cm]{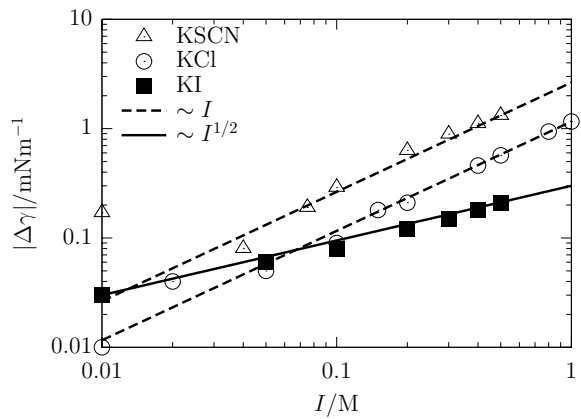}
   \caption{\label{fig:1}Magnitude of the excess interfacial tension $|\Delta\gamma|$ as a function of the
           ionic strength $I$ in water of a water-decaline interface for the three salts $\m{KSCN}$ 
           ($\bigtriangleup$), $\m{KCl}$ ($\bigcirc$), and $\m{KI}$ ($\blacksquare$) according to 
           Ref.~\cite{Guest1939}. The interfacial tension of a salt-free water-decaline interface is 
           $50.94\ \m{mNm^{-1}}$. The dashed lines are power laws $\sim I$ whereas the solid line is a 
           power law $\sim I^{1/2}$ both derived from the present model (see Eq.~\Eq{gammaasymp}).}
\end{figure}

Figure~\ref{fig:1}
displays the magnitude of the excess interfacial tension $|\Delta\gamma|$ of a water-decaline
interface as a function of the ionic strength $I$ in \emph{water} for three different salts, $\m{KSCN}$ 
($\bigtriangleup$), $\m{KCl}$ ($\bigcirc$), and $\m{KI}$ ($\blacksquare$), as published in Ref.~\cite{Guest1939}. 
The interfacial tension of a salt-free water-decaline interface is $50.94\ \m{mNm^{-1}}$. The dashed lines are 
power laws $\sim I$ passing through the largest data
points for $\m{KSCN}$ and $\m{KCl}$, whereas the solid line is a power law $\sim I^{1/2}$ passing through the
smallest data point for $\m{KI}$. Within the present model one concludes from Fig.~\ref{fig:1} that the crossover
ionic strength $I^\times$ for $\m{KSCN}$ and $\m{KCl}$ is smaller than $0.01\m{M}$ whereas for $\m{KI}$ it is 
larger than $0.5\m{M}$. The prediction $\Delta\gamma(I\ll I^\times)<0$ from Eq.~\Eq{gammaasymp} is in agreement
with the data for $\m{KI}$ in Ref.~\cite{Guest1939}. Finally, the excess interfacial tension measured in 
Ref.~\cite{Guest1939} is \emph{negative} for $\m{KSCN}$ and \emph{positive} for $\m{KCl}$. Within the present
model this observation is to be interpreted as follows: For the case of $\m{KSCN}$ ($\Delta\gamma(I \gg I^\times)<0$)
one infers $p>1$ from Eq.~\Eq{gammaasymp}, and consequently in this case solvent $A$ is water and solvent $B$ is 
decaline, because the ionic strength in water is larger than in decaline. Assuming $p\gg 1$ Eq.~\Eq{gammaasymp}
leads to $\Delta\gamma(I\gg I^\times)\simeq -2sI$ with $I=\rho_A$ which, for $\m{KSCN}$, yields
$s \approx 0.53\ \m{nm}$. Hence the discontinuity of the solvent-induced potentials $V_\alpha$ for $\m{KSCN}$ 
is located at a distance $0.53\ \m{nm}$ on the \emph{decaline}-side of a water-decaline 
interface. For $\m{KCl}$ ($\Delta\gamma(I \gg I^\times)>0$), on the other hand, $p<1$ due to Eq.~\Eq{gammaasymp}, 
i.~e., here solvent $A$ is decaline and solvent $B$ is water. Assuming $p\ll 1$ gives rise to 
$\Delta\gamma(I\gg I^\times)\simeq 2sI$ with $I=\rho_B$ which, for $\m{KCl}$, leads to $s\approx 0.23\ \m{nm}$.
Thus the discontinuity of $V_\alpha$ for $\m{KCl}$ is located at a distance $0.23\ \m{nm}$ on the \emph{water}-side
of the water-decaline interface. These findings suggest a weaker affinity of $\m{Cl^-}$ for the organic 
decaline phase than $\m{[SCN]^-}$, which agrees with the structure of these anions. Hence the excess interfacial 
tension data in Ref.~\cite{Guest1939} can be consistently described in terms of Eq.~\Eq{gammaasymp} with respect 
to the sign and the power law in the ionic strength. Moreover, $s$ is, as expected, comparable to the size of the
ions. However, a more detailed experimental check would be highly appreciated.

To conclude, it has been found within a simple model that at small ionic strength $I$ the excess liquid-liquid 
interfacial tension of electrolyte solutions behaves as $\mathcal{O}(-\sqrt{I})$ due to an unequal partitioning of
ions, whereas at large ionic strength it behaves as $\mathcal{O}(\pm I)$ due to a finite interfacial thickness.
These asymptotic regimes are in agreement with the findings of Nichols and Pratt \cite{Nichols1984} and Onuki
\cite{Onuki2006}. The crossover strongly depends on the components of the system such that all 
suggested asymptotic regimes can be realized experimentally by choosing appropriate liquids and electrolytes 
(see Fig.~\ref{fig:1} and Ref.~\cite{Guest1939}). The decrease of the liquid-liquid tension at low $I$ 
is expected to be much more pronounced when highly charged colloids are considered instead of
low-valency ions. Quantitative understandig of this increase is
of direct relevance for the stability of Pickering emulsions as mentioned at the beginning of this work.
On the basis of the present simple model for liquid-liquid interfaces between electrolytic solutions, which in 
this work has proved to agree with experimental data, it should now be possible to study the effect of adding 
surfactants and colloids on the interfacial tension in order to ultimately obtain a fully microscopic theory of 
the (in)stability of emulsions \cite{Aveyard1985,Onuki2006}.

\begin{acknowledgments}
This work is part of the research program of the 'Stichting voor 
Fundamenteel Onderzoek der Materie (FOM)', which is financially supported by 
the 'Nederlandse Organisatie voor Wetenschappelijk Onderzoek (NWO)'. 
\end{acknowledgments}



\end{document}